\documentclass[aps,prx,showpacs,twocolumn,superscriptaddress]{revtex4}
\usepackage{bm,color}
\usepackage[dvipdfmx]{graphicx}
\usepackage{amsmath, amssymb}
\usepackage{braket}
\begin{document}
\title{Machine-Learning Detection of the Berezinskii-Kosterlitz-Thouless Transition and the Second-Order Phase Transition in the XXZ models}
\author{Yusuke Miyajima}
\affiliation
{Department of Applied Physics, Waseda University, Okubo, Shinjuku-ku, Tokyo 169-8555, Japan}
\author{Masahito Mochizuki}
\affiliation
{Department of Applied Physics, Waseda University, Okubo, Shinjuku-ku, Tokyo 169-8555, Japan}
\begin{abstract}
We propose two machine-learning methods based on neural networks, which we respectively call the phase-classification method and the temperature-identification method, for detecting different types of phase transitions in the XXZ models without prior knowledge of their critical temperatures. The XXZ models have exchange couplings which are anisotropic in the spin space where the strength is represented by a parameter $\Delta(>0)$. The models exhibit the second-order phase transition when $\Delta>1$, whereas the Berezinskii-Kosterlitz-Thouless (BKT) phase transition when $\Delta<1$. In the phase-classification method, the neural network is trained using spin or vortex configurations of well-known classical spin models other than the XXZ models, e.g., the Ising models and the XY models, to classify those of the XXZ models to corresponding phases. We demonstrate that the trained neural network successfully detects the phase transitions for both $\Delta>1$ and $\Delta<1$, and the evaluated critical temperatures coincide well with those evaluated by conventional numerical calculations. In the temperature-identification method, on the other hand, the neural network is trained so as to identify temperatures at which the input spin or vortex configurations are generated by the Monte Carlo thermalization. The critical temperatures are evaluated by analyzing the optimized weight matrix, which coincide with the result of numerical calculation for the second-order phase transition in the Ising-like XXZ model with $\Delta=1.05$ but cannot be determined uniquely for the BKT transition in the XY-like XXZ model with $\Delta=0.95$.
\end{abstract}

\maketitle
\section{INTRODUCTION}
Machine learning techniques are capable of solving classification and regression problems by extracting hidden features in data~\cite{LeCun15,Goodfellow16}. Recently, the techniques have begun to be exploited for scientific research~\cite{Carleo19}. It is widely recognized that the architecture of machine learning that performs classification, prediction, and presumption by learning certain features from enormous data has compatibility with the procedure of scientific research particularly in physics, which approaches universal concepts of nature by extracting essences from accumulated data and knowledge.

Indeed, the machine learning techniques have recently been applied to the condensed-matter physics intensively~\cite{Bedolla21}. One important subject in research of this kind is description of quantum many-body states using neural networks~\cite{Carleo17,Nomura17,Nomura21}, which is based on the universal approximation theorem of neural networks and the variational ansatz. Recently, the ground-state phase diagram of a frustrated $J_1$-$J_2$ quantum Heisenberg model has been revealed by calculating the energies for optimized neural networks describing quantum spin states~\cite{Nomura21}. This phase diagram has not been obtained by any numerical methods so far. In this sense, this is an important example that a machine learning elucidated a new physics.

Another important subject of the machine-learning research in condensed-matter physics is detection of phases and phase transitions in physical models~\cite{Ohtsuki20}. Recently, phase diagrams of several quantum models with randomness have been investigated using machine learning techniques by classifying spatial-map images of electron wavefunctions at various parameters to corresponding phases. The machine learning detections of phases and phase transitions are basically based on image recognition and pattern classification. Several kinds of classical spin models, in particular, the Ising models in two dimensions, have been investigated as research targets for benchmark tests~\cite{Tanaka17,Carrasquilla17,Arai18,Suchsland18,Hu17,Wetzel17,Wang16,Ponte17,Nieuwenburg17,Liu18,Giannetti19,Bachtis20a}. The second-order phase transition accompanied by spontaneous symmetry breaking in the square-lattice Ising model exhibits apparent difference in the spatial spin configurations between the paramagnetic phase and the (anti)ferromagnetic phase because of a trivial change in the order parameter. Therefore, the machine learning technique can detect the phase transition in this model rather easily.

According to the Mermin-Wagner's theorem~\cite{Mermin66}, spin models with continuous symmetry do not show any spontaneous symmetry breaking at finite temperatures in one and two dimensions. Indeed, the XY model in two dimensions has the U(1) symmetry and thus does not exhibit any phase transitions with symmetry breaking. However, this model is known to exhibit the BKT transition which is a topologically characterized phase transition between a high-temperature phase with individual vortices and antivortices and a low-temperature phase with vortex-antivortex bound pairs~\cite{Kosterlitz73,Kosterlitz74}. It is known that the BKT transition is difficult to detect by numerical techniques based on the statistical mechanics such as the Monte Carlo techniques. This is because anomalies, i.e., jump, peak, and kink, in the thermodynamic quantities are not necessarily related with the BKT transition point directly. For example, a peak of the specific heat does not show up at the transition point. The magnetic susceptibility exhibits an anomaly at the transition point but is suffered from significant finite-size effects with a logarithmic correction term, which makes a size-scaling extrapolation to the thermodynamic limit difficult.

Recently, several attempts have been made to detect the BKT transition in the XY model and the $q$-state clock model with machine learning techniques~\cite{Richter-Laskowska18,Beach18,ZhangW19,Rodriguez-Nieva19,Tran21,Wang21,Mendes-Santos21,Singh21,Haldar22,Shiina20,Tomita20,Miyajima21}. However, because the BKT transition is a phase transition without any spontaneous symmetry breakings, the machine learning techniques based basically on the pattern recognition are not necessarily powerful for its detection. Most of the attempts are based on supervised learning methods, and they require feature engineering in advance to make preprocessed input data, that is, the data of quantity which captures the features of the phases and the phase transitions such as vortex configurations, histogram of the spin orientations, and the spin correlation functions. Moreover, they need prior knowledge of fundamental properties of the model, e.g., number of phases, approximate transition temperatures, and order parameters. These are critical problems because the first one means that we must know features that characterize the phases and the phase transitions hosted by the models in advance, and the second one means that the method cannot be applied to unknown models.

Here it should also be mentioned that there have been a lot of experimental efforts to search for the BKT behavior in real magnetic materials so far~\cite{ZHu2020,Tutsch14,Kohama11,Cuccoli03a,Cuccoli03b}, but they are not necessarily successful. A main obstacle for the observation of the BKT behavior in real materials is the small but almost unavoidable interlayer coupling combined with the smallness of the easy-plane anisotropy. From the theoretical point of view, the lack of powerful theoretical framework to discuss possible BKT behavior in complicated spin models might be another obstacle. The BKT phases and the BKT transitions in theoretical spin models have been studied usually by using the Monte Carlo techniques. In the Monte Carlo analysis, the quantity called helicity modulus is usually employed to detect the BKT transition. The expression of this quantity does not have a general form but has a form specific to each spin model, which must be derived by hand for each case. Its derivation becomes difficult for complicated theoretical spin models that describe real magnets with further exchange coupling, magnetic anisotropies, and higher-order interactions. Therefore, a powerful and versatile machine-learning method applicable to general spin models is highly demanded for further development of the research on the BKT behavior in real magnets.

In this paper, we propose two machine-learning methods based on neural networks with simple architectures as versatile tools to detect the BKT phase and the BKT transition in classical spin models. The two methods are referred to as the phase-classification method and the temperature-identification method, respectively. We employ the XXZ models for examination of these methods and try to detect two types of phase transitions, i.e., the second-order transition and the BKT transition, in the XXZ models without prior knowledge of their critical temperatures. The XXZ models have exchange couplings which are anisotropic in the spin space with an anisotropy parameter $\Delta(>0)$ and exhibits the second-order transition when $\Delta>1$ (Ising-like case) and the BKT transition when $\Delta<1$ (XY-like case). In the phase-classification method, the spin or vortex configurations of the XXZ models generated by the Monte Carlo thermalization are classified into corresponding phases by utilizing a neural network trained with those of well-known classical spin models, e.g., the Ising model and the XY model as inspired by previous studies in Refs.~\cite{Kim18,Bachtis20b,Fukushima21}. We demonstrate that the trained neural network successfully performs the classification and determines the phase-transition point for both $\Delta>1$ and $\Delta<1$. The evaluated critical temperatures coincide well with those evaluated by numerical calculations. In the temperature-identification method, on the other hand, the neural network is trained so as to output correct temperatures at which the input spin or vortex configurations are generated by the Monte Carlo thermalization, and the critical temperatures are evaluated by analyzing the optimized weight matrix after the training as inspired by studies in Refs.~\cite{Tanaka17,Arai18}. We show that the critical temperature evaluated by this analysis coincides with the result of Monte Carlo calculation for the second-order phase transition for $\Delta=1.05$, whereas the critical temperature cannot be evaluated uniquely for the BKT transition for $\Delta=0.95$. 

Both of our proposed methods have advantages over the Monte Carlo methods that have been traditionally used for studying the phase-transition phenomena in spin models in terms of computational cost and generality. Moreover, our methods also have advantages over the previously proposed machine learning methods, that is, our methods require no or less prior knowledge of the models and no or least featrure engineering with preprocessing of input data in contrast to the previous methods. These advantages will be discussed in the last section in detail. We expect that our work will pave the way to exploring novel phases and phase-transition phenomena in theoretical spin models and will contribute to the further development of this research field.

\section{MODEL}
We study ferromagnetic ($J>0$) XXZ models on square lattices. The Hamiltonian is given by,
\begin{equation}
\label{Eq01}
\mathcal{H}=
-J\sum_{\langle i,j \rangle} \left(S_i^x S_j^x + S_i^y S_j^y + \Delta S_i^z S_j^z \right),
\end{equation}
where $\bm{S}_i = \left(S_i^x, S_i^y, S_i^z \right)$ is a continuous classical spin vector on the $i$th site, norm of which is set to be unity ($|\bm{S}_i|=1$). Here the parameter $\Delta (>0)$ describes the anisotropy of the exchange coupling in the spin space, and the summation is taken over the nearest-neighbor site pairs $\langle i,j \rangle$. The XXZ models have been studied intensively because many materials and systems such as ultrathin ferromagnetic films and superconducting films have turned out to be well described by these models~\cite{Hikami80,Kawabata86,Serena93,Cuccoli95,Pires96,Lee05,Aoyama19,Shirinyan19}.

\begin{figure}
\includegraphics[scale=1.0]{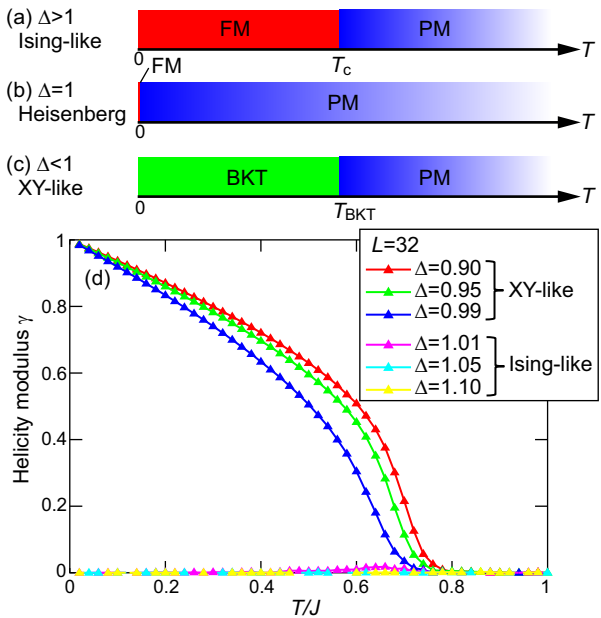}
\caption{Schematic temperature phase diagrams of the XXZ models on a square lattice for (a) $\Delta>1$ (Ising-like case), (b) $\Delta=1$ (Heisenberg case), and (c) $\Delta<1$ (XY-like case). Here FM and PM denote the ferromagnetic and paramagnetic phases, respectively. (d) Temperature profiles of the helicity modulus calculated using the Monte Carlo technique for various values of $\Delta$. This quantity becomes finite in the XY-like case with $\Delta<1$ below a certain temperature, whereas it vanishes in the Ising-like case with $\Delta>1$. The calculations are performed for a system size of $L^2$ with $L=32$.}
\label{Fig01}
\end{figure}
Schematic temperature phase diagrams of the XXZ models are shown in Figs.~\ref{Fig01}(a)-(c). We find that the models exhibit distinct behaviors depending on the anisotropy $\Delta$. When $\Delta>1$, the model exhibits a single second-order phase transition as in the Ising model, whereas the model exhibits a single BKT transition as in the XY model when $\Delta<1$. At $\Delta=1$, the model is equivalent to the classical Heisenberg model in two dimensions and thus does not exhibit any phase transition at finite temperatures.

To see the $\Delta$-dependent behaviors of the XXZ models, we calculate a quantity called helicity modulus~\cite{Fisher73,Nelson77,Weber88} by measuring variations of the free energy when uniform and infinitesimal twists are introduced in the spin alignment in a certain direction. Figure~\ref{Fig01}(d) shows temperature profiles of the helicity modulus $\gamma$ for various values of the anisotropy parameter $\Delta$ calculated by the Monte Carlo calculations. The calculations are performed for a square lattice of $32 \times 32$ sites, and the twist along the $x$-axis is considered. We find that $\gamma$ is suppressed to be zero at any temperatures for the Ising-like case with $\Delta>1$, whereas it increases with decreasing temperature below a certain temperature for the XY-like case with $\Delta<1$ which manifests the emergence of BKT phase.

\section{METHOD}
\begin{figure*}
\includegraphics[scale=1.0]{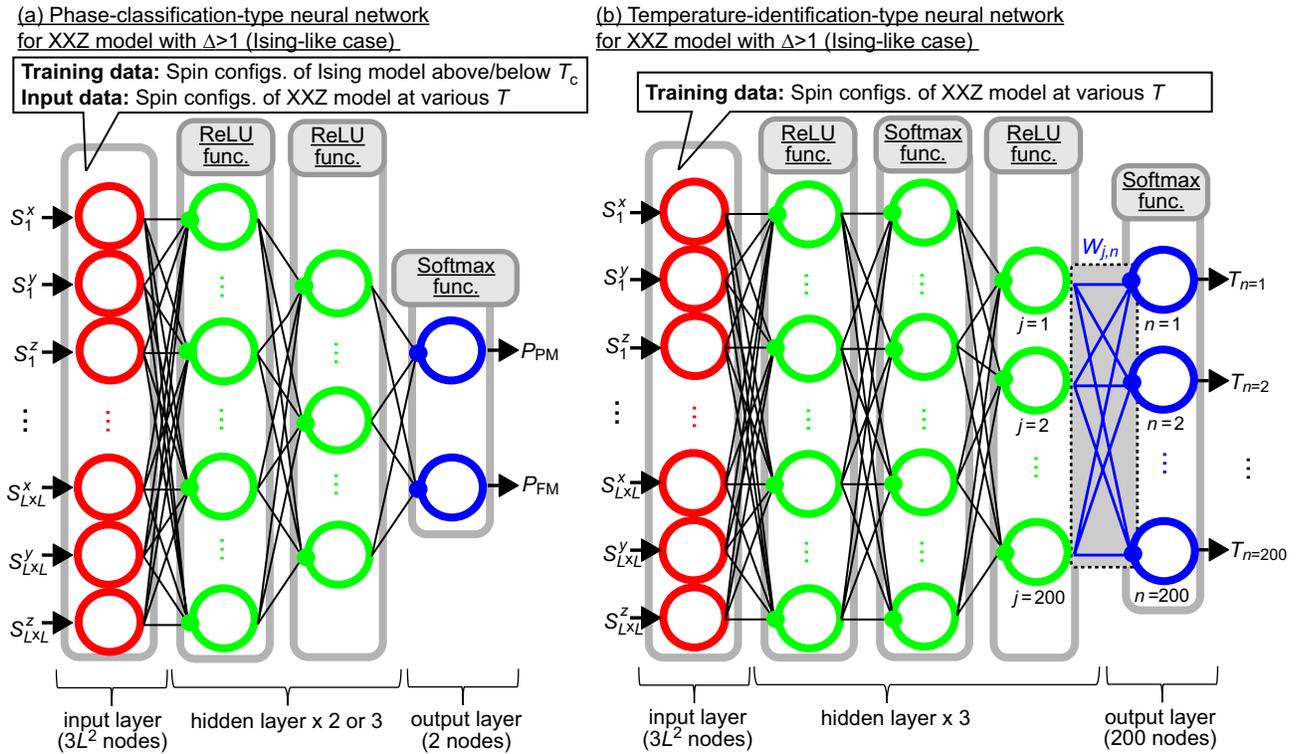}
\caption{Neural networks used for detection of the second-order phase transition in the Ising-like XXZ model with $\Delta>1$. (a) Phase-classification type neural network. Spin configurations of $\bm S_i=(0, 0, S_{iz})$ with $S_{iz}=+1$ or $-1$ for the Ising model are used for training data, while those of $\bm S_i=(S_{ix},S_{iy},S_{iz})$ for the XXZ model are used for the input data. This neural network output probabilities $P_{\rm PM}$ and $P_{\rm FM}$ that the input spin configuration is paramagnetic and ferromagnetic, respectively, where their sum $P_{\rm PM}+P_{\rm FM}$ is normalized to be unity. The weight matrices connecting the layers in this neural network are trained so as to output a higher probability for the phase to which the spin configuration indeed belongs. (b) Temperature-identification-type neural network.  Spin configurations of $\bm S_i=(S_{ix},S_{iy},S_{iz})$ for the XXZ model are used for training data. The weight matrices in this neural network are trained so as to output a correct temperature at which the input spin configuration is obtained. All the spin configurations used as training and input data are prepared by the Monte Carlo thermalization. The activation functions adopted in respective hidden and output layers are also presented.}
\label{Fig02}
\end{figure*}
\begin{figure*}
\includegraphics[scale=1.0]{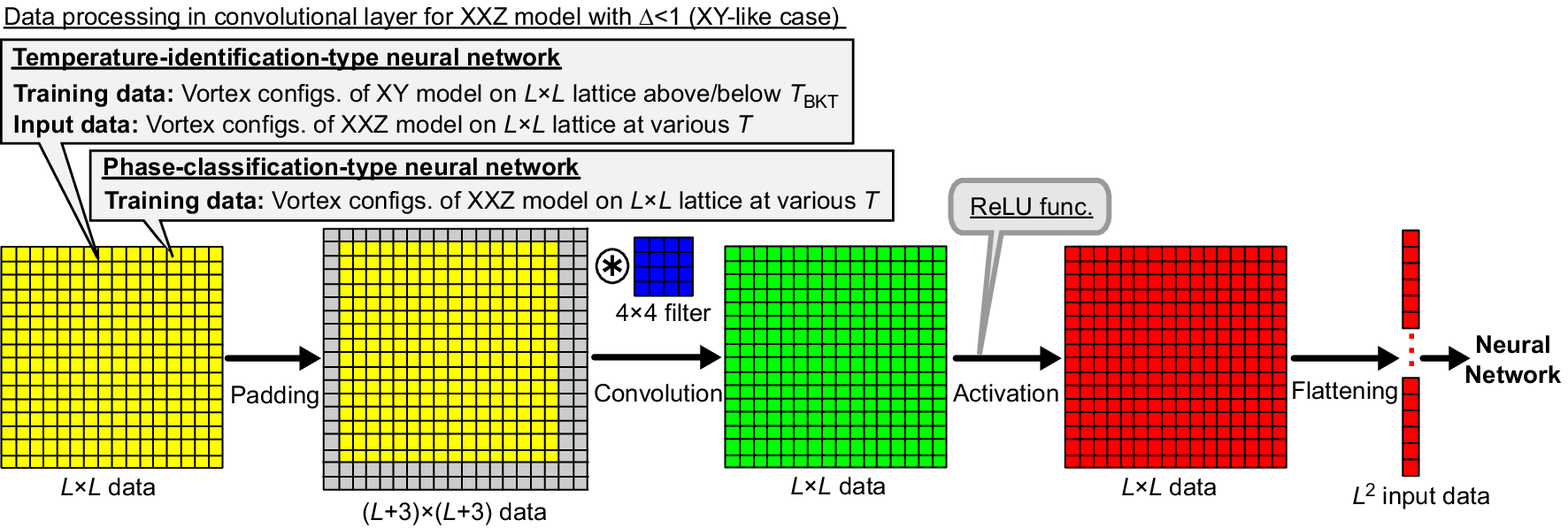}
\caption{Data convolution process in the convolutional neural network for detection of the BKT transition in the XY-like XXZ model with $\Delta<1$ on the $L \times L$ lattice. Spatial vortex-configuration data with $L^2$ components are made from the spin-configuration data with $3L^2$ components generated by the Monte Carlo thermalization. The whole process of the convolution layer contains four procedures, i.e., padding, convolution, activation, and flattening. The $L^2$ data are transformed to $(L+3)^2$ data by adding zeros as additional components in the padding procedure. A $4 \times 4$ filter is applied to the $(L+3)^2$ data to obtain $L^2$ convoluted data in the convolution procedure, while the ReLU function is adopted for the activation procedure. The processed $L^2$ data are fed to a neural network, e.g., the phase-classification-type neural network or the temperature-identification-type neural network, which are basically the same as those in Fig.~\ref{Fig04}(a) and (b), respectively, although the number of input nodes is different. For the phase-classification-type neural network, the vortex configurations for the XY model are used as training data, while those of the XXZ model is used as input data. This neural network output probabilities $P_{\rm PM}$ and $P_{\rm BKT}$ that the input vortex configuration belongs to the paramagnetic phase and the BKT phase, respectively, where their sum $P_{\rm PM}+P_{\rm BKT}$ is normalized to be unity. The weight matrix in this neural network as well as $4^2$ components of the filter are trained so as to assign the inputted vortex configurations to correct phases. For the temperature-identification-type neural network, the vortex configurations of the XXZ model are used to train the weight matrix and the filter so as to output correct temperatures at which the input vortex configurations are generated by the Monte Carlo thermalization.}
\label{Fig03}
\end{figure*}
We examine two machine-learning methods based on neural networks for detecting the phase transitions in the XXZ models, which are referred to as the phase-classification method and the temperature-identification method, respectively. The neural networks used in these methods are implemented with the machine-learning library KERAS~\cite{KERAS15}. We perform supervised learning for these neural networks using a set of training input and corresponding answer labels.

\subsection{Phase-classification method}
We first implement a neural network shown in Fig.~\ref{Fig02}(a). The binary classification of two phases is considered here. Inputs to this neural network are spin or vortex configurations of well-known classical spin models other than the XXZ models, i.e., the Ising models and the XY models in the present study, prepared by the Monte Carlo thermalization. The outputs are vectors with two components, each of which represents a probability that the input belongs to the phase. This neural network is referred to as the phase-classification type hereafter.

For detection of the second-order phase transition in the Ising-like XXZ model with $\Delta=1.05$, we simply use this neural network, which comprises one input layer, one or two hidden layers, and one output layer. The input layer has $3L^2$ nodes corresponding to all the spin components on the $L \times L$ lattice, whereas the output layer has two nodes corresponding to the ferromagnetic phase and the paramagnetic phase. The two outputs are probabilities $P_{\rm FM}(T)$ and $P_{\rm PM}(T)$ that the input belongs to the ferromagnetic phase and the paramagnetic phase, respectively. The number of the hidden nodes in each hidden layer is determined so as to encode the $3L^2$-component inputs into the two-component outputs with the same compression ratio at each layer. Adjacent layers are connected by weight matrices which weight the importance of data transmitted from the former layer to the next layer. The Rectified Linear Unit (ReLU) function is employed as an activation function for the hidden layers, whereas the softmax function for the output layer. The sum of probabilities $P_{\rm FM}(T)+P_{\rm PM}(T)$ equals to unity because the softmax function normalizes the components of output. 

For detection of the BKT transition in the XY-like XXZ model with $\Delta=0.95$, the input data are processed by using the convolutional neural network~\cite{Goodfellow16} before feeding them into the phase-classification-type neural network. The convolutional neural network is adopted to capture two-dimensional features of spatial configurations of vortices and antivortices to judge whether they are bound or unbound. The convolutional neural network comprises one input layer, three hidden layers, and one output layer as shown in Fig.~\ref{Fig03}, where the first hidden layer is the convolution layer. The ReLU function is employed for the three hidden layers, whereas the softmax function for the output layer. In the convolution layer, the input data with $L^2$ components are transformed into a data with $(L+3)^2$ components by adding zeros as additional components in the so-called zero-padding procedure. The convolutional procedure is performed by applying a filter with $4^2$ weight components to the $(L+3)^2$ data to obtain $L^2$ data. Then the ReLU function is applied to each component of the processed $L^2$ data for activation, and the two-dimensional data are flattened to one-dimensional $L^2$ data.

The convoluted $L^2$ data are fed to the phase-classification-type neural network for detection of the BKT transition. The two nodes in the output layer correspond to the BKT phase and the paramagnetic phase, which output the probabilities $P_{\rm BKT}(T)$ and $P_{\rm PM}(T)$ that the input vortex configuration belongs to the BKT phase and the paramagnetic phase, respectively. The sum of probabilities $P_{\rm BKT}(T)+P_{\rm PM}(T)$ again equals to unity. We determine the number of nodes in the hidden layers in the same manner as the Ising-like case.

We prepare sets of the spin (vortex) configurations of the Ising (XY) model and corresponding phase labels as training data. The neural network is trained so as to correctly guess phases to which the input spin (vortex) configurations belong. We use a cost function to quantify the difference between the outputs of the neural network and the answer labels. As a cost function, the cross-entropy error function is employed, which is given by,
\begin{equation}
\label{Eq02}
E(\bm{t}^i, \bm{y}^i) = \sum_i \sum_k t_k^i \log y_k^i,
\end{equation}
where $i$ is the label of input data, and $k$ is the index of output nodes. $y_k^i$ represents the $k$th component of the output vector from the neural network, and $t_k^i$ represents the $k$th component of the answer label for the $i$th input training data. In the training procedure, we minimize the loss function by tuning the components of weight matrices using the optimization algorithm called Adam~\cite{Kingma14}.

After the training procedure, we feed 100 spin or vortex configurations of the XXZ model generated at each of 200 temperature points to the neural network as the input data. For the $l$th input data ($l=1,2, \cdots, 100$) generated at temperature $T$, the probabilities $P_1^l(T)$ and $P_2^l(T)$ are output from the first and the second nodes in the output layer, respectively. From the 100 outputs of $P_\alpha^l(T)$ ($\alpha=1, 2$), we calculate averaged probabilities $\bar{P}_\alpha(T)$ at each temperature point as,
\begin{equation}
\label{Eq03}
\bar{P}_\alpha(T)=\frac{1}{N_l} \sum_{l=1}^{N_l}P_\alpha^l(T)
\quad\quad (N_l=100).
\end{equation}
At the critical temperature, the neural network is confused in the phase classification because pronounced thermal-fluctuation effect obscures the difference between two phases. Thus, we determine the critical temperature by the crossing point of $\bar{P}_1(T)$ and $\bar{P}_2(T)$, at which both $\bar{P}_1(T)$ and $\bar{P}_2(T)$ take 0.5.

\subsection{Temperature-identification method}
We next implement a neural network shown in Fig.~\ref{Fig02}(b). The multiclass classification of temperatures is carried out by this neural network. Inputs to this neural network are spin or vortex configurations of the XXZ model prepared by the Monte Carlo thermalization at 200 temperature points ranged from $T_{n=1}=0.01J$ to $T_{n=200}=2.00J$ with constant intervals of $\Delta T=0.01J$. The outputs are vectors with 200 components, where the $n$th component represents a probability that the input spin or vortex configuration is generated at $T_n=n\Delta T$. This neural network is referred to as the temperature-identification type hereafter.

For detection of the second-order phase transition in the Ising-like XXZ model with $\Delta=1.05$, we simply use this neural network, which comprises one input layer, three hidden layers, and one output layer. The input layer has $3L^2$ nodes corresponding to all the spin components on the $L \times L$ lattice, whereas the output layer has 200 nodes corresponding to the 200 temperature points. The number of the hidden nodes in each hidden layer is determined so as to encode the $3L^2$-component inputs into 200-component outputs with the same compression ratio at each layer. Adjacent layers are connected by weight matrices. The ReLU function is employed as an activation function for the first and third hidden layers, while the softmax function is employed for the second hidden layer and the output layer.

On the other hand, we use the temperature-identification-type convolutional neural network~\cite{Goodfellow16} for detecting the BKT transition in the XY-like XXZ model with $\Delta=0.95$. The convolutional neural network comprises one input layer, three hidden layers, and one output layer, where the first hidden layer is the convolution layer. As activation functions, the ReLU function is employed for the first and the third hidden layer, the softmax function is employed for the second hidden layer and the output layer. The same data processing as in the convolution layer of phase-classification-type convolutional neural network is performed in the convolution layer. In the convolutional neural network, input data is processed as follows: First, two-dimensional input data is transformed into one-dimensional data in the convolution layer. The convolution layer connects the second hidden layer in Fig.~\ref{Fig02}(b). The rest processes are the same as the neural network in Fig.~\ref{Fig02}(b). Here, the output layer has 200 nodes corresponding to the 200 temperature points, while the number of the hidden nodes is determined as the same as $\Delta=1.05$. 

We prepare sets of spin configurations of the Ising-like XXZ model with $\Delta=1.05$ or vortex configurations of the XY-like XXZ model with $\Delta=0.95$ and corresponding temperature labels as training data. Therefore, the neural networks are trained to correctly estimate the temperature at which the input spin or vortex configurations are generated. We use the cross-entropy error function as a cost function. In the training process, using the optimization algorithm called Adam~\cite{Kingma14}, we minimize the loss function by tuning components of weight matrices. 

After the training procedure, we focus on the weight matrix $W_{j,n}$ connecting the $j$th node in the last hidden layer with $j=1, 2, \cdots 200$ and the $n$th node in the output layer corresponding to the $n$th temperature point $T_n=n\Delta T$ with $n=1, 2, \cdots 200$. A gray-scale plot of the matrix components $W_{j,n}$ in plane of $j$ and $T_n$ is called heat map. Heat maps tend to exhibit distinct patterns for different phases, which can be exploited to detect phase transitions.

We evaluate critical temperatures by quantitative analysis of the pattern changes in heat maps. For this purpose, we introduce special kinds of correlation function $C_W(T)$ and variance $V_W(T)$. The correlation function $C_W(T)$ is defined by,
\begin{equation}
\label{Eq04}
C_W(T_n)=\frac{1}{N_{\rm h}}\sum_{m=1}^{n-1} \sum_{j=1}^{N_{\rm h}} W_{j,m}W_{j,m+1},
\end{equation}
where $N_h(=200)$ is the number of nodes in the last hidden layer. The product $W_{j,m}W_{j,m+1}$ quantifies similarity or difference between adjacent components. On the other hand, the variance $V_W(T)$ is defined by,
\begin{equation}
\label{Eq05}
\begin{split}
& V_W(T_n)=\frac{1}{n-1}\sum_{j=1}^{N_h}\sum_{m=1}^{n-1}(W_{j,m} - \bar{W}_{j})^2, \\
& \bar{W}_j=\frac{1}{n-1} \sum_{m=1}^{n-1}W_{j,m}.
\end{split}
\end{equation}
For $T_n<T_{\rm c}$, heat maps tend to exhibit a stripe pattern reflecting an ordered phase. Consequently, the components $W_{j,n}$ take similar values within each row specified by $j$ and thus the variance $V_W(T_n)$ is small. On the contrary, for $T_n>T_{\rm c}$, heat maps tend to exhibit a sandstorm pattern reflecting a disordered phase, and the components $W_{j,n}$ take scattered values along each row. As a result, the variance $V_W(T_n)$ increases from $T_{\rm c}$ with increasing $T_n$. Pattern changes in the heat map manifests themselves in the $T_n$-profiles of $C_W(T_n)$ and $V_W(T_n)$ in distinct manners. 

\section{Results for $\Delta=1.05$ (Ising-like case)}
\subsection{Monte Carlo calculation}
\begin{figure}
\includegraphics[scale=1.0]{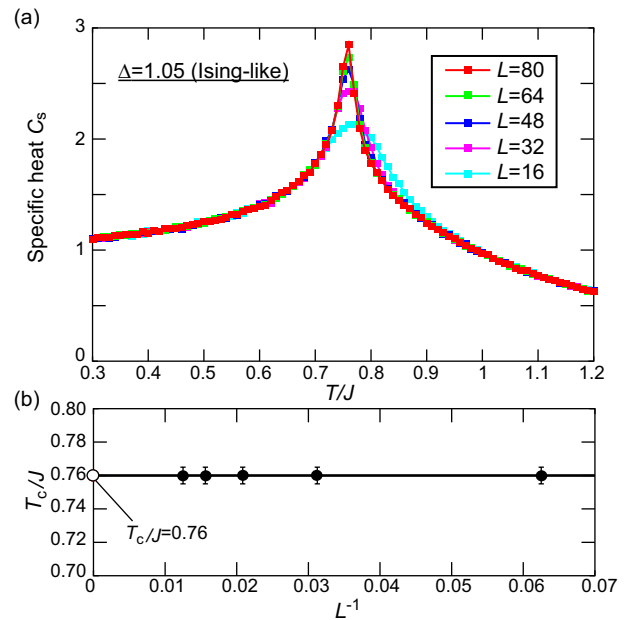}
\caption{(a) Temperature profiles of specific heat for the Ising-like XXZ model with $\Delta=1.05$ for various system sizes, each of which exhibits a peak at the critical temperature $T_{\rm c}$ of the second-order phase transition from the paramagnetic to ferromagnetic phases. (b) Finite-size scaling analysis of $T_{\rm c}$. In the thermodynamic limit of $L \rightarrow \infty$, $T_{\rm c}$ is evaluated to be $T_{\rm c}/J=0.76$ by extrapolation.}
\label{Fig04}
\end{figure}
\begin{table}
\caption{Critical temperatures of the Ising-like XXZ model with $\Delta=1.05$ for various system sizes in units of $J$. The case of $L\rightarrow \infty$ denote results in the thermodynamic limit.}
\begin{tabular}{ccccccc} 
\hline \hline
$L$ & 16 & 32 & 48 & 64 & 80 & $\infty$ \\
\hline 
$T_{\rm c}^{\rm MC}/J$ & 0.760(5) & 0.760(5) & 0.760(5) & 0.760(5) & 0.760(5) & 0.760(0) \\
$T_{\rm c}^{\rm PC}/J$ & 0.724(6) & 0.735(3) & 0.749(4) & 0.755(3) & 0.763(6) & 0.766(5) \\
\hline \hline
\end{tabular}
\label{Tab01}
\end{table}
We first determine the critical temperature $T_{\rm c}$ of the second-order phase transition in the Ising-like XXZ model with $\Delta=1.05$ from specific heat calculated by the Monte Carlo calculation. Figure~\ref{Fig04}(a) shows the calculated temperature profiles of specific heat for several lattice sizes. For each lattice size, the specific heat has a peak at $T_{\rm c}$. A finite-size scaling analysis is performed to evaluate $T_{\rm c}$ in the thermodynamic limit. We derive the size dependence of $T_{\rm c}$. The spin correlation length $\xi$ behaves as $\xi \propto |T-T_{\rm c}|$ near the critical temperature because the second-order phase transition belongs to the Ising universality class. Because the correlation length $\xi$ satisfies $\xi=L$ at the critical temperature $T_{\rm c}(L)$ for finite-size systems, the size dependence of the critical temperature is given by, 
\begin{equation}
\label{Eq06}
T_{\rm c}(L)=T_{\rm c}(\infty)+\frac{a}{L}.
\end{equation}
Here $T_{\rm c}(\infty)$ is the critical temperature in the thermodynamic limit, and $a$ is a constant. Figure~\ref{Fig04}(b) presents the size dependence of the critical temperature $T_{\rm c}(L)/J$. From the extrapolation to $L^{-1}\rightarrow 0$, the critical temperature in the thermodynamic limit is evaluated as $T_{\rm c}^{\rm MC}/J=0.76$. The critical temperatures for several system sizes are summarized in Table~\ref{Tab01}.

\subsection{Spin configurations used as training data}
\begin{figure}[t]
\includegraphics[scale=1.0]{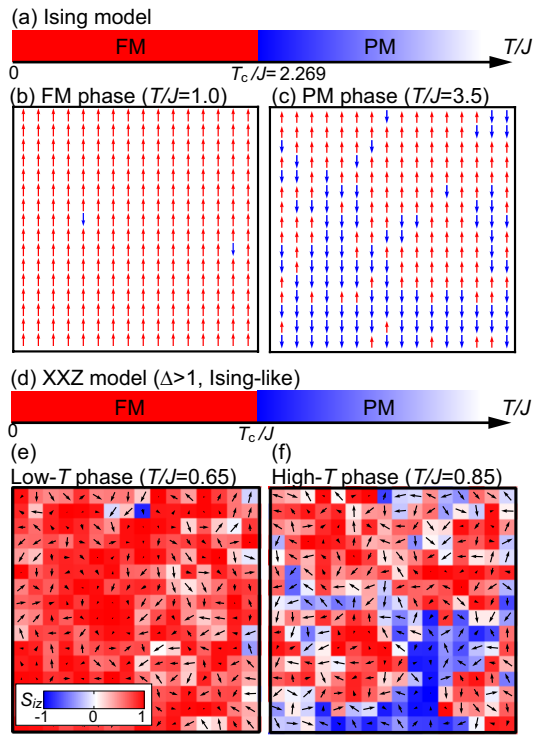}
\caption{Spin configurations of the Ising model on square lattices are used for data to train the neural network aiming for detection of the second-order phase transition in the XXZ model with $\Delta>1$. (a) Schematic temperature phase diagram of the square-lattice Ising model, which exhibits the second-order phase transition from the paramagnetic (PM) phase to the ferromagnetic (FM) phase at $T_{\rm c}/J=2.269$. Spin configurations generated by the Monte Carlo thermalization in the temperature range $0.01 \le T/J \le 2.01$ ($2.52 \le T/J \le 4.00$) are used for the training data for the FM (PM) phase. (b), (c) Typical spin configurations of (b) FM phase ($T/J=1.0$, $L=16$) and (c) PM phase ($T/J=3.5$, $L=16$). Spin configurations of the XXZ model with $\Delta=1.05$ at various temperatures are entered to the trained neural network as input data. (d) Schematic temperature phase diagram of the Ising-like XXZ model with $\Delta>1$. (e), (f) Typical spin configurations at (e) lower temperatures ($T/J=0.65$) and (f) higher temperatures ($T/J=0.85$) of the XXZ model with $\Delta = 1.05$. It turns out that the neural network trained by the binary-spin configurations of the Ising model is capable of detecting the phase transition of the three-dimensional vector-spin states in the XXZ model.}
\label{Fig05}
\end{figure}
We prepare many spin configurations of the Ising model and the Ising-like XXZ model with $\Delta=1.05$ for several lattice sizes $L \times L$ ($L=16, 32, 48, 64, 80$) as training data of the neural network. The spin configurations are generated by the Monte Carlo thermalization with 12000 iterative steps based on the single-flip Metropolis algorithm. For the Ising model, 100 spin configurations are prepared at each of 400 temperature points from $T=0.01J$ to $T=4.00J$ at intervals of $\Delta T=0.01J$. For the Ising-like XXZ model, on the other hand, 200 spin configurations are prepared at each of 200 temperature points from $T=0.01J$ to $T=2.00J$ at intervals of $\Delta T =0.01J$. 

The temperature phase diagram and the examples of spin configurations for the Ising model are presented in Figs.~\ref{Fig05}(a)-(c). The blue (red) arrows in Figs.~\ref{Fig05}(b) and (c) denote $+1$ $(-1)$ spins. Most of the spins are pointing in the same direction in the ferromagnetic phase [Fig.~\ref{Fig05}(b)], whereas $+1$ and $-1$ spins are randomly distributed in the paramagnetic phase [Fig.~\ref{Fig05}(c)]. On the other hand, a schematic temperature phase diagram and the examples of spin configurations for the Ising-like XXZ model are presented in Figs.~\ref{Fig05}(d)-(f). The arrows in Figs.~\ref{Fig05}(e) and (f) denote the in-plane components of the spins, while the colors of pixels represent their out-of-plane components. The image for the ferromagnetic phase [Fig.~\ref{Fig05}(e)] is almost unicolored by red or blue because most of the ordered spins are pointing along the $z$ axis, while the red and blue pixels are randomly distributed in the image for the paramagnetic phase [Fig.~\ref{Fig05}(f)]. The existence of in-plane spin components might make the situation complicated for detection of the phase transition in the XXZ model.

\subsection{Phase-classification method}
\begin{figure}
\includegraphics[scale=1.0]{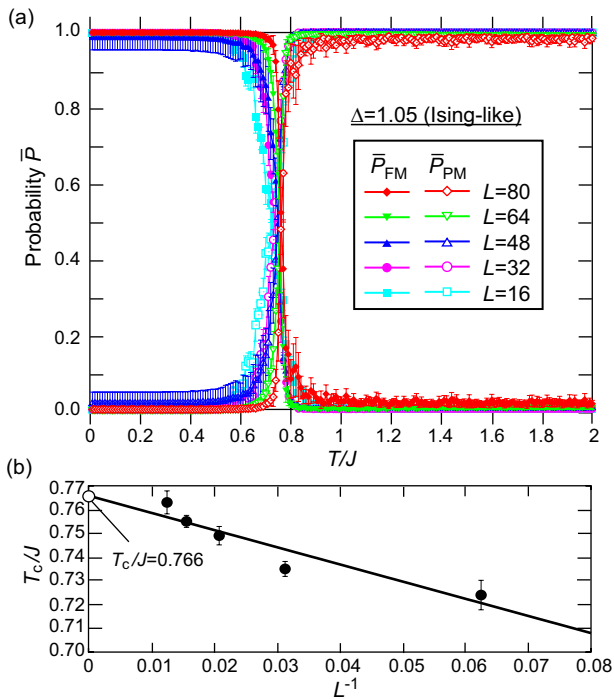}
\caption{(a) Temperature profiles of the averaged probabilities $\bar{P}_{\rm PM}$ and $\bar{P}_{\rm FM}$ of the Ising-like XXZ model with $\Delta=1.05$ obtained by the phase-classification-type neural network for various system sizes. Here, $\bar{P}_{\rm PM}(T)$ and $\bar{P}_{\rm FM}(T)$ are averages of the probabilities that a set of the spin configurations generated at temperature $T$ by the Monte Carlo thermalization are classified to the paramagnetic phase and the ferromagnetic phase, respectively. The summation $\bar{P}_{\rm PM}+\bar{P}_{\rm FM}$ is always normalized to be unity. Each crossing point of $\bar{P}_{\rm PM}$ and $\bar{P}_{\rm FM}$, at which both probabilities take 0.5, is assigned to a transition temperature for the corresponding system size. (b) Finite-size scaling analysis of $T_{\rm c}$. $T_{\rm c}$ in the thermodynamic limit of $L \rightarrow \infty$ is evaluated to be $T_{\rm c}/J=0.766$ by extrapolation, which coincides well with $T_{\rm c}/J=0.76$ evaluated by the conventional Monte Carlo analysis.}
\label{Fig06}
\end{figure}
We attempt to detect the second-order phase transition in the Ising-like XXZ model with $\Delta=1.05$ by the phase-classification-type neural network shown in Fig.~\ref{Fig02}(a). As the training dataset, 100 ferromagnetic spin configurations of the Ising model (instead of the XXZ model) are prepared at each temperature point ranged from $T=0.01$ to $T=2.01$, while 100 paramagnetic spin configurations are prepared within a temperature range from $T=2.52$ to $T=4.00$ at intervals of $\Delta T=0.01$. Here, the spin configurations around $T_{\rm c}=2.269$ are excluded from the training data. Each spin configuration of the Ising model on the $L \times L$ square lattice has only $L^2$ components. We, therefore, transform the binary spin variables $S_i(=\pm1)$ into three-component vectors $(0, 0, S_i)$. This transformation is justified by the uniaxial magnetic anisotropy in the XXZ models. 

The vectors $(0, 0, S_1, \cdots, 0, 0, S_i, \cdots, 0, 0, S_{L^2})$, which is arrays of the vectors $(0, 0, S_i)$ at all sites, is fed to the neural network. On the other hand, the answer labels for the training data are vectors with two components, which represent probabilities that the input spin configuration belongs to the paramagnetic phase and the ferromagnetic phase, respectively. The vector $(1, 0)$ is given as the answer labels for the input spin configurations generated below $T_{\rm c}$, while the vector $(0, 1)$ is given for those generated above $T_{\rm c}$. The former (latter) vector mean that the input is absolutely for the ferromagnetic (paramagnetic) phase. The neural network is trained so as to correctly guess phases to which the input spin configurations belong.

After completing the training procedure, we input the spin configurations of the XXZ model to the trained neural network, where each input is an array of the spin vectors, i.e., $(S_1^x, S_1^y, S_1^z, \cdots, S_i^x, S_i^y, S_i^z, \cdots, S_{L^2}^x, S_{L^2}^y, S_{L^2}^z)$. As the iput dataset, 100 spin configurations are prepared at each temperature point within a range from $T=0.01J$ to $T=2.00J$ at intervals of $\Delta T=0.01J$. For the 100 spin configurations generated at $T$, 100 output vectors ($P_{\rm FM}^l(T)$, $P_{\rm PM}^l(T)$) are obtained, where $l$ ($=1, 2, \cdots, 100$) is the index of the input data. We calculate the averages of $P_{\rm FM}^l(T)$ and $P_{\rm PM}^l(T)$ over 100 input data to obtain the averaged probabilities $\bar{P}_{\rm FM}(T)$ and $\bar{P}_{\rm PM}(T)$.

Figure~\ref{Fig05}(a) shows the temperature profiles of $\bar{P}_{\rm FM}(T)$ and $\bar{P}_{\rm PM}(T)$. We find that $\bar{P}_{\rm FM}(T) \approx 1$ and $\bar{P}_{\rm PM}(T) \approx 0$ at lower temperatures, whereas $\bar{P}_{\rm FM}(T) \approx 0$ and $\bar{P}_{\rm PM}(T) \approx 1$ at higher temperatures. This indicates that the neural network correctly guesses phases to which the input data belong. The critical temperature $T_{\rm c}$ for each lattice size is determined from a crossing point of the two profiles. Moreover, $T_{\rm c}$ in the thermodynamic limit is evaluated to be $T_{\rm c}^{\rm PC}/J=0.766$ by the finite-size scaling analysis using Eq.~(\ref{Eq06}). This value coincides well with the value $T_{\rm c}^{\rm MC}/J=0.76$ obtained by the Monte Carlo calculation. The values of $T_{\rm c}$ for several lattice sizes evaluated by the present phase-classification method are summarized in Table~\ref{Tab01}. Now it has been successfully demonstrated that the phase-classification-type neural network can detect the phases and the second-order phase transition in the Ising-like XXZ model.

\subsection{Temperature-identification method}
\begin{figure}
\includegraphics[scale=1.0]{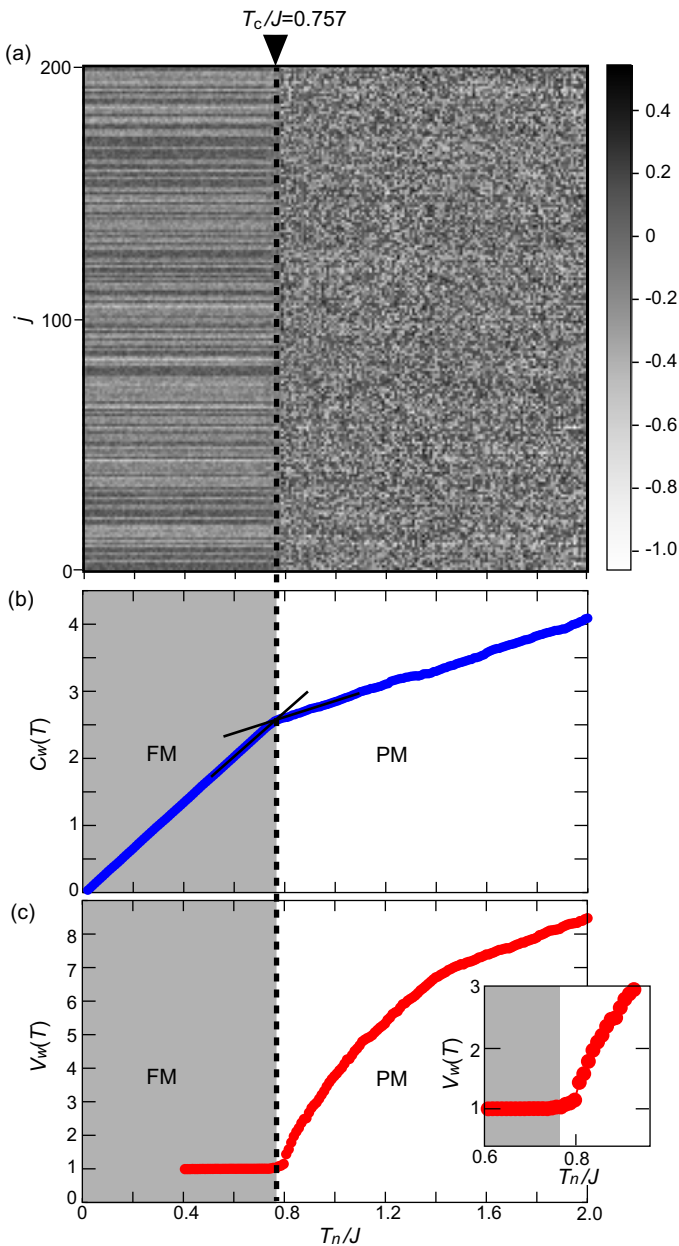}
\caption{(a) Heat map of the weight-matrix components $W_{j,n}$ in the temperature-identification-type neural network, which connect the $j$th node in the last hidden layer and the $n$th node in the output layer in the plane of $j$ ($=1, 2, \cdots, 200$) and $T_n=n\Delta T$ ($n=1, 2, \cdots, 200$) for the Ising-like XXZ model with $\Delta=1.05$ on the square lattice of $L=80$. A clear change in pattern is seen at $T_{\rm c}/J=0.757$, which is ascribed to the phase transition from the paramagnetic (PM) phase to the ferromagnetic (FM) phase at this temperature. (b) Correlation function $C_W(T)$ of the weight-matrix components $W_{j,n}$ defined by Eq.~(\ref{Eq03}) as a function of $T_n$, which clearly shows a change in slope at $T_{\rm c}$. (c) Variance $V_W(T)$ of the weight-matrix components $W_{j,n}$ defined by Eq.~(\ref{Eq04}) as a function of $T_n$, which shows an abrupt change at $T_{\rm c}$.}
\label{Fig07}
\end{figure}
Next we attempt to detect the second-order phase transition in the Ising-like XXZ model with $\Delta=1.05$ by the temperature-identification-type neural network shown in Fig.~\ref{Fig02}(b). As the training dataset, 100 spin configurations of the XXZ model are prepared at each temperature point $T_n=n\Delta T$ ($n=1, 2, \cdots, 200$) ranged from $T=0.01J$ to $T=2.00J$ at intervals of $\Delta T=0.01J$. The vectors $(S_1^x, S_1^y, S_1^z, \cdots S_i^x, S_i^y, S_i^z, \cdots, S_{L^2}^x, S_{L^2}^y, S_{L^2}^z)$, which are arrays of the spin vectors $(S_i^x, S_i^y, S_i^z)$ at all sites $i$ ($=1, 2,\cdots , L^2$), are fed to the neural network as inputs for training. As the answer labels, we adopt temperatures $T_n$, at which the input spin configurations are generated. Here, the answer label is represented by a vector with 200 components. When the answer is $T/J=T_n/J$, only the $n$th component of the vector is set to be unity, while the other components are set to be zero. The neural network is trained so as to correctly guess temperatures at which the input spin configurations are generated.

After the training procedure, we focus on the weight matrix connecting the last hidden layer and the output layer. Components of the weight matrix are visualized as a heat map in Fig.~\ref{Fig07}(a). Here, the horizontal axis represents the index $n$ of output nodes or corresponding temperature $T_n/J(=n\Delta T/J)$, and the vertical axis represents the index $j$ of hidden nodes. A clear change in pattern is observed between the regimes above and below $T \approx 0.76$, where a stripe (sandstorm) pattern appears in the lower (higher) temperature regime. To analyze this pattern change more quantitatively, we calculate the correlation function $C_W(T)$ and variance $V_W(T)$, which are, respectively, defined by Eq.~(\ref{Eq04}) and Eq.~(\ref{Eq05}). Figures~\ref{Fig07}(b) and (c) show the temperature profiles of $C_W(T)$ and $V_W(T)$, respectively. The plot of $C_W(T)$ shows a clear change in slope associated with the phase transition at $T_{\rm c}/J=0.757$. The plot of $V_W(T)$ also shows a clear difference in behavior between the regimes above and below $T_{\rm c}$. It is constant below $T_{\rm c}$, starts increasing at $T_{\rm c}$, and monotonically increases above $T_{\rm c}$. The evaluated critical temperature coincides well with that obtained by the conventional Monte Carlo analysis.

Here it should be mentioned that the variance $V_W(T_n)$ is calculated using the weight-matrix components $W_{j,m}$ at discretized temperature points $T_m$ ($m=1,2,\cdots, n-1$) below $T_n$. When $T_n$ is small, the number of the temperature points that can be used for the variance calculation, i.e., $n-1$, is small so that the calculated variance $V_W(T_n)$ might contain a certain error. Therefore, the variances $V_W(T_n)$ in Fig.~\ref{Fig07}(c) are calculated only for $T_n/J>0.4$ to suppress the errors. The calculated profile of $V_W(T_n)$ is very smooth, and the errors are negligibly small even around $T_n/J \sim 0.4$. We think that this variance-based analysis is applicable even to much lower temperatures.

\section{Results for $\Delta=0.95$ (XY-like case)}
\subsection{Monte Carlo calculation}
\begin{figure}
\includegraphics[scale=1.0]{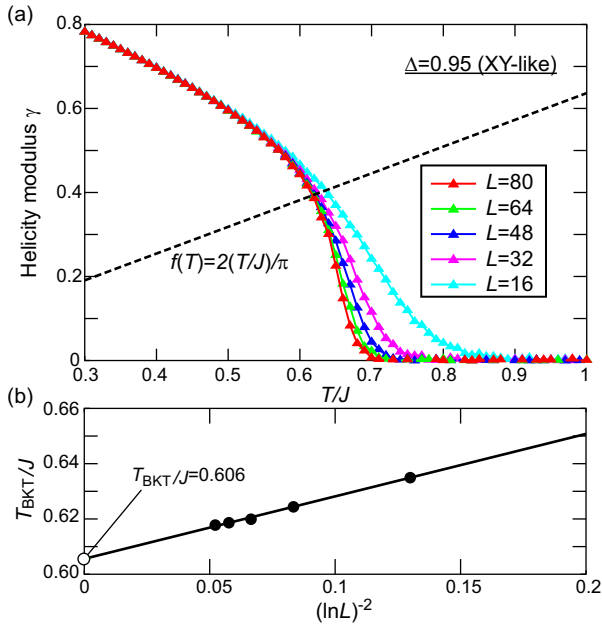}
\caption{(a) Temperature profiles of helicity modulus for the XY-like XXZ model with $\Delta=0.95$ for various system sizes. A point at which the curve and the dashed line cross is assigned to the transition temperature $T_{\rm BKT}$ from the paramagnetic phase to the BKT phase for each system size. The dashed line represents a linear function $f(T)=2(T/J)/\pi$. (b) Finite-size scaling analysis of $T_{\rm BKT}$. In the thermodynamic limit of $L \rightarrow \infty$, $T_{\rm BKT}$ is evaluated to be $T_{\rm BKT}/J=0.606$ by extrapolation.}
\label{Fig08}
\end{figure}
\begin{table*}
\caption{Critical temperatures of the XY-like XXZ model with $\Delta=0.95$ for various system sizes. The case of $L\rightarrow \infty$ denote results in the thermodynamic limit.}
\begin{tabular}{ccccccccc} 
\hline \hline
$L$ & 16 & 32 & 48 & 64 & 80 & 96 & 112 & $\infty$ \\
\hline 
$T_{\rm BKT}^{\rm MC}/J$ & 0.6351 & 0.6243 & 0.6201 & 0.6188 & 0.6177 & - & - & 0.606 \\
$T_{\rm BKT}^{\rm PC}/J$ & - & - & 0.574(5) & 0.576(2) & 0.578(5) & 0.582(9) & 0.584(7) & 0.603(5) \\
\hline \hline
\end{tabular}
\label{Tab02}
\end{table*}
We determine the critical temperature $T_{\rm BKT}$ of the BKT transition in the XY-like XXZ model from helicity modulus $\gamma$ calculated by the Monte Carlo calculation. The helicity modulus of the XXZ model is calculated using the following formula given in the form of thermal average,
\begin{multline}
\label{Eq07}
\gamma_{\rm XXZ}
=\frac{J}{L^2}
\left\langle
\sum_{\langle i,j \rangle_x}
\sqrt{1-S_{iz}^2} \sqrt{1-S_{jz}^2} \cos(\phi_i - \phi_j)
\right\rangle \\
-\frac{\beta J^2}{L^2}
\left\langle
\left(\sum_{\langle i,j\rangle_x} 
\sqrt{1-S_{iz}^2} \sqrt{1-S_{jz}^2} \sin(\phi_i - \phi_j) 
\right)^2 \right \rangle.
\end{multline}
For the derivation of this formula, see Appendix. This quantity exhibits a universal jump from $\gamma=0$ to $\gamma=2T_{\rm BKT}/\pi$ at $T_{\rm BKT}$. For each lattice size, $T_{\rm BKT}$ is determined by the crossing point between the temperature profiles of $\gamma$ and the line $f(T)=2T/\pi$. Fig.~\ref{Fig08}(a) shows the temperature profiles of $\gamma_{\rm XXZ}$ for the XY-like XXZ model with $\Delta=0.95$ for several lattice sizes.

We evaluate the BKT transition temperature in the thermodynamic limit $T_{\rm BKT}(\infty)$ by a finite-size scaling analysis. Kosterlitz and Thouless revealed that the correlation length behaves as $\exp(c/\sqrt{|t|})$ at temperatures slightly above $T_{\rm BKT}(\infty)$. Here $c$ is a constant, and $t$ is a relative temperature defined by,
\begin{equation}
\label{Eq08}
t=\frac{T-T_{\rm BKT}(\infty)}{T_{\rm BKT}(\infty)}.
\end{equation}
For finite-size systems, $\xi$ equals to the lattice size $L$ at $T_{\rm BKT}$. Therefore, $T_{\rm BKT}(L)$ is given as a function of the system size $L$,
\begin{equation}
\label{Eq09}
T_{\rm BKT}(L)=T_{\rm BKT}(\infty) + \frac{c^\prime}{(\ln L)^2}.
\end{equation}
The size dependence of $T_{\rm BKT}(L)$ for the XY-like XXZ model with $\Delta=0.95$ is shown in Fig.~\ref{Fig04}(b). We evaluate $T_{\rm BKT}^{\rm MC}/J=0.606$ in the thermodynamic limit from the extrapolation. The obtained values of $T_{\rm BKT}(L)$ for several sizes are summarized in Table~\ref{Tab02}.

\subsection{Vortex configurations used as training data}
\begin{figure}
\includegraphics[scale=1.0]{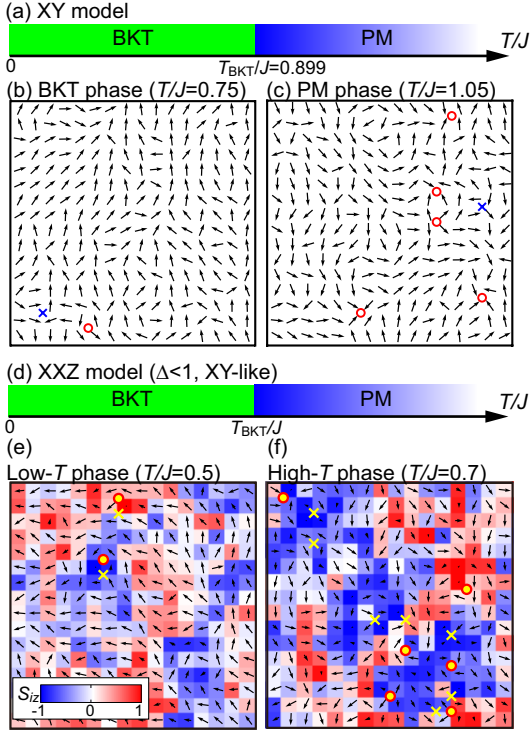}
\caption{Vortex configurations of the XY models on square lattices are used for data to train the neural network aiming for detection of the BKT transition in the XXZ model with $\Delta<1$. (a) Schematic temperature phase diagram of the square-lattice XY model, which exhibits the BKT transition from the paramagnetic (PM) phase to the BKT phase at $T_{\rm BKT}/J=0.899$. Vortex configurations generated by the Monte Carlo thermalization in the temperature range $0.75 \lesssim T/J \le 0.89$ ($0.90 \le T/J \lesssim 1.05$) are used for the training data for the BKT (PM) phase. (b), (c) Typical spin and vortex configurations of (b) BKT phase ($T/J=0.75$, $L=16$) and (c) PM phase ($T/J=1.05$, $L=16$). Locations of vortices and antivortices are indicated by solid circles and cross symbols. Spin configurations of the XXZ model with $\Delta = 0.95$ at various temperatures are entered to the trained neural network as input data. (d) Schematic temperature phase diagram of the XY-like XXZ model with $\Delta<1$. (e), (f) Typical spin configurations at (e) lower temperatures ($T/J=0.5$) and (f) higher temperatures ($T/J=0.7$) of the XXZ model with $\Delta=0.95$. It turns out that the neural network trained by the two-dimensional planar-spin configurations of the XY model can detect the phase transition of the three-dimensional vector-spin states in the XXZ model.}
\label{Fig09}
\end{figure}
We prepare many vortex configurations of the XY model and the XY-like XXZ model with $\Delta=0.95$ for several lattice sizes $L \times L$ ($L=16, 32, 48, 64, 80, 96, 112$) as training data of the neural network, which are made from the spin configurations by calculating the vorticity on each plaquette. Here the spin configurations are generated by the Monte Carlo thermalization with $10^5$ iterative steps based on the single-flip Metropolis algorithm. For the XY model and the XY-like XXZ model, vortex configurations are prepared at each temperature point.

The temperature phase diagram and the examples of vortex configurations for the XY model are presented in Figs.~\ref{Fig09}(a)-(c). The red circles and blue crosses indicate vortices and antivortices, respectively, in Figs.~\ref{Fig09}(b) and (c). Corresponding spin configurations are also shown in these figures where the arrows represent the in-plane components of spin vectors. One vortex-antivortex bound pair is seen in Fig.~\ref{Fig09}(b) for the BKT phase, whereas six free vortices and antivortex are seen in Fig.~\ref{Fig09}(c) for the paramagnetic phase. On the other hand, a schematic temperature phase diagram and the examples of vortex configurations for the XY-like XXZ model are presented in Figs.~\ref{Fig09}(d)-(f). Two vortex-antivortex bound pairs are seen in Fig.~\ref{Fig09}(e) for the BKT phase, whereas there are many free vortices and antivortices in Fig.~\ref{Fig09}(f) for the paramagnetic phase. Here the colors of pixels represent their out-of-plane components or the $z$ components. 

Here it should be mentioned that we have also examined the bare spin configurations as input data instead of the vortex configurations. We have found that the machine learning method often fails to detect the BKT transition in this case. The neural network tends to detect the growth of global magnetization component more sensitively than the evolution of vortex configuration, which makes it difficult to detect the BKT transition characterized by the bound and unbound states of vortex-antivortex pairs. In fact, this aspect has been pointed out in previous work~\cite{Beach18}. This problem might be more crucial for the weakly easy-plane XXZ models with three-dimensional continuous vector spins than the XY model with pure planar vector spins. In our previous work~\cite{Miyajima21}, we examined the $q$-state clock model and found that the machine-learning method can detect the BKT transition successfully even with bare spin configurations as input data, which might be attributable to the two-dimensional discrete spin degrees of freedom.

\subsection{Phase-classification method}
\begin{figure}
\includegraphics[scale=1.0]{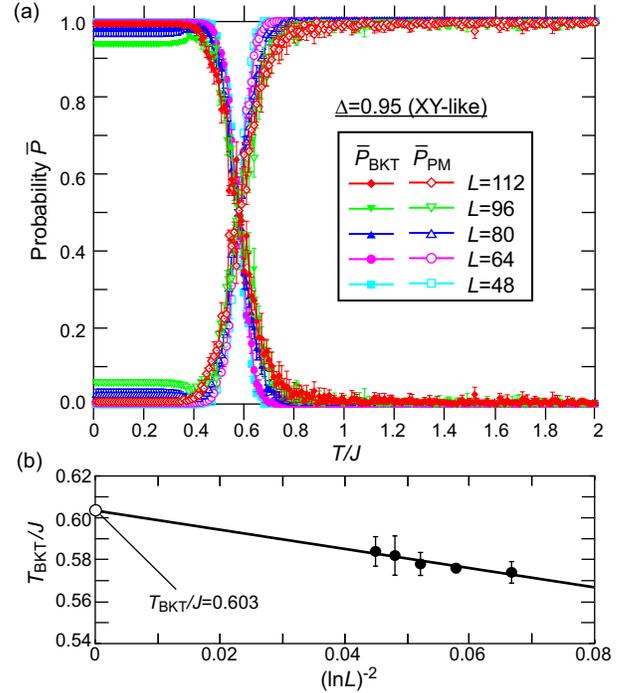}
\caption{(a) Temperature profiles of the averaged probabilities $\bar{P}_{\rm PM}$ and $\bar{P}_{\rm BKT}$ of the XY-like XXZ model with $\Delta=0.95$ obtained by the phase-classification-type neural network for various system sizes, where the summation $\bar{P}_{\rm PM}+\bar{P}_{\rm BKT}$ is always normalized to be unity. Each crossing point of $\bar{P}_{\rm PM}$ and $\bar{P}_{\rm BKT}$ is assigned to a transition temperature for the corresponding system size. (b) Finite-size scaling analysis of $T_{\rm BKT}$. $T_{\rm BKT}$ in the thermodynamic limit of $L \rightarrow \infty$ is evaluated to be $T_{\rm BKT}/J=0.603$ by extrapolation, which coincides well with $T_{\rm BKT}/J=0.606$ evaluated by the conventional Monte Carlo analysis.}
\label{Fig10}
\end{figure}
We attempt to detect the BKT phase transition in the XY-like XXZ model with $\Delta=0.95$ by the phase-classification-type neural network [Fig.~\ref{Fig02}(a)] combined with the convolutional neural network [Fig.~\ref{Fig03}]. As the training dataset, 100 vortex configurations for the BKT (paramagnetic) phase in the XY model (instead of the XXZ model) are prepared at each temperature point within a range from $T \sim 0.75J$ to $T=0.89J$ (from $T=0.90J$ to $T \sim 1.05J$) at intervals of $\Delta T=0.01J$.

We input vectors with $L^2$ components, which are arrays of the vorticities $V_i$ on all plaquettes in the square lattice with periodic boundary conditions, to the neural network. On the other hand, the answer labels for the training data are vectors with two components, which represent probabilities that the input vortex configuration belongs to the BKT phase and the paramagnetic phase, respectively. More specifically, $(1, 0)$ is given as answer labels for the input vortex configurations prepared at $T<T_{\rm BKT}$, whereas $(0, 1)$ is given for those prepared at $T>T_{\rm BKT}$. The neural network is trained so as to correctly guess phases to which the input vortex configurations belong.

After completing the training procedure, we input the vortex configurations of the XXZ model to the trained neural network, where each input is an array of the $L^2$ vorticities. As the input dataset, 100 vortex configurations are prepared at each temperature point within a range from $T=0.01J$ to $T=2.00J$ at intervals of $\Delta T=0.01J$.For the 100 vortex configurations generated at $T$, 100 output vectors ($P_{\rm BKT}^l(T)$, $P_{\rm PM}^l(T)$) are obtained, where $l$ ($=1,2, \cdots, 100$) is the index of the input data. We calculate the averages of $P_{\rm BKT}^l(T)$ and $P_{\rm PM}^l(T)$ over 100 input data to obtain the averaged probabilities $\bar{P}_{\rm BKT}(T)$ and $\bar{P}_{\rm PM}(T)$.

Figure~\ref{Fig10}(a) shows the temperature profiles of $\bar{P}_{\rm BKT}(T)$ and $\bar{P}_{\rm PM}(T)$. In these plots, $\bar{P}_{\rm BKT}(T) \approx 1$ and $\bar{P}_{\rm PM}(T) \approx 0$ at lower temperatures, whereas $\bar{P}_{\rm BKT}(T) \approx 0$ and $\bar{P}_{\rm PM}(T) \approx 1$ at higher temperatures, indicating  that the neural network correctly guess phases to which the input data belong. The critical temperature $T_{\rm BKT}$ for each lattice size is determined from a crossing point of the two profiles. Moreover, $T_{\rm BKT}$ in the thermodynamic limit is evaluated as $T_{\rm BKT}^{\rm PC}/J=0.603$ by the finite-size scaling analysis using Eq.~(\ref{Eq09}). This value coincides well with the value $T_{\rm BKT}^{\rm MC}/J=0.606$ obtained by the Monte Carlo calculation. The values of $T_{\rm BKT}$ for several lattice sizes evaluated by the present phase-classification method are summarized in Table~\ref{Tab02}. Now we have succeeded in demonstrating the detections of the BKT transition in the XY-like XXZ model by the phase-classification-type neural network.

\subsection{Temperature-identification method}
\begin{figure}
\includegraphics[scale=1.0]{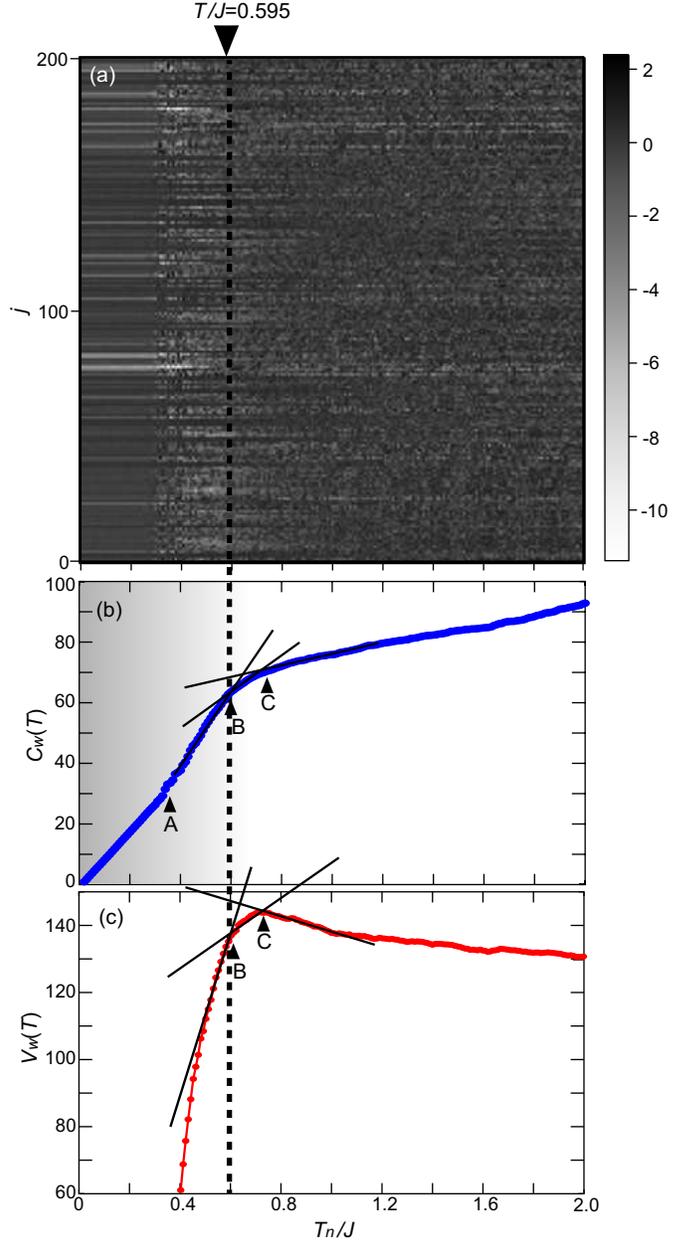}
\caption{(a) Heat map of the weight-matrix components $W_{j,n}$ in the temperature-identification-type neural network for the XY-like XXZ model with $\Delta=0.95$ on the square lattice of $L=80$. (b), (c) Temperature profiles of (b) correlation function $C_W(T)$ and (c) variance $V_W(T)$ of the weight-matrix components $W_{j,n}$. These profiles exhibit changes in slope at several points (A, B, and C), but A and C do not correspond to the BKT transition points. On the other hand, B is located almost at the real transition point of $T_{\rm BKT}/J=0.606$.}
\label{Fig11}
\end{figure}
\begin{figure}
\includegraphics[scale=1.0]{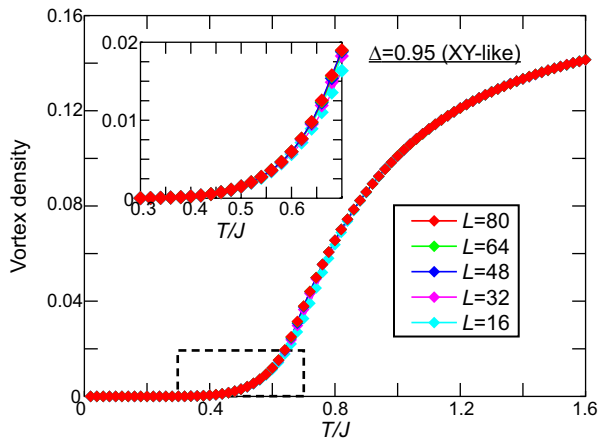}
\caption{Temperature profiles of the vortex density in the XY-like XXZ model with $\Delta=0.95$ on square lattices for various system sizes. Inset magnifies an area indicated by the dashed rectangle, around which the vortex density starts increasing.}
\label{Fig12}
\end{figure}
Finally we attempt to detect the BKT transition in the XY-like XXZ model with $\Delta=0.95$ by the temperature-identification-type neural network combined with the convolutional neural network. As the training dataset, 100 vortex configurations of the XXZ model are prepared at each temperature point ranged from $T=0.01J$ to $T=2.00J$ at intervals of $\Delta T=0.01J$. We input the vortex configurations to the neural network in the form of vectors with $L^2$ components. As the answer labels, we adopt temperatures $T_n$, at which the input vortex configurations are generated. Again the neural network is trained so as to correctly guess temperatures at which the input vortex configurations are generated.

After the training procedure, we again focus on the weight matrix of the neural network connecting the last hidden layer and the output layer, components of which are visualized as a heat map in Fig.~\ref{Fig11}(a). We find that a stripe (sandstorm) pattern appears in the lower (higher) temperature regime. However, their boundary is difficult to identify by eyes because the stripe and sandstorm patterns are mixed in the in-between regime of $0.3<T_n/J<0.7$. To analyze the pattern change more quantitatively, we calculate the correlation function $C_W(T)$ and variance $V_W(T)$ defined by Eq.~(\ref{Eq04}) and Eq.~(\ref{Eq05}). Again we calculated $V_W(T_n)$ for $T_n/J>0.4$.

Figs.~\ref{Fig11}(b) and (c) present the temperature profiles of $C_W(T)$ and $V_W(T)$, respectively. We find that $C_W(T)$ exhibits changes in slope three times at points A, B, and C, whereas $V_W(T)$ exhibits abrupt changes in behavior twice at B and C. To understand possible origins of these anomalies, we calculate temperature profiles of the vortex density for several lattice sizes of $L \times L$ with $L=16, 32, 48, 64, 80$ (Fig.~\ref{Fig12}). Here, the vortex density is calculated as the total number of vortices and antivortices per one plaquette. In these profiles, we find that A corresponds to a temperature at which the profiles start to increase, whereas C corresponds to its inflection point. At the point A, the neural network might detect the emergence of vortices and antivortices. On the other hand, the anomaly C might be related with a peak of the specific heat. We indeed observed the specific-heat peak around this temperature in the Monte Carlo calculations (not shown). Because the specific heat and the vortex density are related with each other through the spin configurations, they could have anomalies (i.e., a peak and an inflection) simultaneously when a certain crossover occurs. The neural network might detect an anomaly associated with the crossover around the point C. On the contrary, the anomaly B seems to correspond to $T_{\rm BKT}$. Accordingly, we conclude that the temperature-identification-type convolutional neural network seemingly captures the BKT transition in the XY-like XXZ model, but it fails to determine the transition point uniquely.

\section{Conclusion and Discussion}
In summary, we have examined two machine learning methods, which are, respectively, referred to as the phase-classification method and the temperature-identification method, for detecting the second-order transition in the Ising-like XXZ model with a weak easy-axis anisotropy of $\Delta=1.05$ and the BKT transition in the XY-like XXZ model with a weak easy-plane anisotropy $\Delta=0.95$.

In the phase-classification method, the neural network trained by the spin configurations of the Ising model is used to classify the spin configurations of the Ising-like XXZ model into the ferromagnetic phase and the paramagnetic phase. On the other hand, the neural network trained by the vortex configurations of the XY model is used to classify the vortex configurations of the XY-like XXZ model into the BKT phase and the paramagnetic phase. For both cases, we have evaluated the critical temperatures for several system sizes and have performed finite-size scaling analyses. The obtained critical temperatures in the thermodynamic limit coincide well with those evaluated in the previous studies for both cases.

In the temperature-identification method, the neural network is trained so as to correctly answer the temperatures at which the spin (vortex) configurations of the XXZ model with $\Delta=1.05$ ($\Delta=0.95$) are generated by the Monte Carlo thermalization. After the training is completed, we focus on the weight matrix connecting the last hidden layer and the output layer to evaluate the critical temperatures. We have introduced two quatities, i.e., the correlation function $C_W(T)$ and the variance $V_W(T)$ to quantitatively analyze the pattern changes in the heat map that visualizes the weight-matrix components. The evaluated critical temperature for the second-order transition in the XXZ model with $\Delta=1.05$ coincide with that obtained in the previous study. On the contrary, in the case of the XXZ model with $\Delta=0.95$, the neural network detects not only the BKT transition but also other cross-over behaviors and thus fails to determine the BKT transition temperature uniquely.

The phase transitions in theoretical spin models have been usually studied by statistical analyses based on the Monte Carlo technique. Our machine learning methods have some advantages over this traditional technique. First, the computational cost can be significantly reduced. Both of our machine learning methods require the spin or vortex configurations generated by the Monte Carlo thermalization, but they do not necessarily have to be thermal equilibrium states. Namely even imperfectly thermalized spin or vortex configurations can work for the detection of the phase transitions, which enables us to save the numerical cost significantly. This is in striking contrast to the Monte Carlo method, which requires the huge number of samplings after sufficient thermalization procedure to improve the accuracy of the thermal averages. We have confirmed that the analysis which takes approximately an hour by our machine learning methods typically takes several hours or even a few days if it is done by the Monte Carlo method.

Second, our machine learning methods do not require the derivation of model-sensitive expression of the helicity modulus for detection of the BKT transition. As mentioned in the introduction section, the physical quantity called helicity modulus is usually used to detect the BKT transition in the Monte Carlo analysis. The expression of this quantity does not have a general form but has a form specific to each spin model, which must be derived by hand for each case. Its derivation becomes difficult when the theoretical spin model is not simple. Our methods can avoid this difficulty and thus can be applied to general spin models. In addition, in the case of the phase-classification method, the neural network can be applied to other models without any extra training procedure, if it is once trained.

We should also mention that several machine learning methods have been applied to theoretical spin models as argued in the introduction section, most of which are based on supervised learning techniques. Our proposals have several advantages over them. One advantage is that our methods require no or less information of the model in advance. Conventional machine learning methods often require the prior knowledge of fundamental properties of the model, e.g, the number of phases, the approximate transition temperature(s), and even the order parameter(s) to assign the answer labels to the training data. Thus, they include a contradiction that the properties of the model we are about to examine must be known in advance. In contrast, both of our methods require the prior knowledge of neither the transition temperatures nor the order parameters. In the case of the temperature-identification method, we do not need to know even the number of phases.

For detection of phase transitions in theoretical spin models, the previously proposed machine learning methods often require preprocessing of input data to extract the features of the phases and phase transitions. In particular, for detection of the BKT phase, the methods using the vortex configurations, the spin-direction histogram, and some correlation functions have been proposed, which are made from the bare spin-configuration data. Among these quantities, the vortex configurations are relatively natural to be employed because the BKT behavior is characterized by the bound and unbound states of the vortex-antivortex pairs. Moreover, they are easily made from the spin-configuration data. In this sense, the least requirement of the feature engineering is another advantage of our methods. Here we note that in Ref.~\cite{Miyajima21}, we applied the temperature-identification method to the $q$-state clock model with $q=8$ and demonstrated that the BKT phase and the successive BKT transitions in this model can be successfully detected even using the bare spin configurations as input data. This might be attributed to the fact that less degrees of freedom of the two-dimensional discretized spin vectors of the clock models. On the contrary, the three-dimensional continuous spin vectors of the weakly easy-plane XXZ models have larger degrees of freedom, which might make the detection of the BKT behavior more difficult.

Finally, we discuss possible future challenges for further development of our work. First, the examination of the XXZ models with different anisotropy values might be worth doing. In the present work, we have examined only two cases with $\Delta=1.05$ and $\Delta=0.95$, which are, respectively, the Ising-like case with a weak easy-axis anisotropy and the XY-like case with a weak easy-plane anisotropy on the verge of their boundary. We chose these rather difficult cases to test the efficiency of our methods. However, it must be interesting to examine how the performances of our machine learning methods depend on the distance from the boundary at $\Delta=1$, which is expected to scale with the clarity of phase transitions. Second, the application of the phase-classification method to other spin models that can host the BKT phase is also an issue of interest. Through this research, we might be able to clarify what features contained in the input data are indeed detected by the neural network and to establish a guiding principle of the systematic machine learning scheme for detection of phase transitions in spin models.

\begin{widetext}
\section{Appendix A: Helicity modulus for the XXZ model}
The helicity modulus is a physical quantity which describes the degree of increase in free energy when a global twist is applied to the system. Specifically, in a system with lattice size of $L$ in one direction, we consider the gradual twisting of spins with an equivalent twisting angle of $\delta/L$ from the left end to the right end so as to achieve a situation that the rightmost spin is twisted by an infinitesimal angle $\delta$ relative to the leftmost spin. When the Hamiltonian contains only terms symmetric with respect to the spin exchange, the increment of the free energy $F$ in this case does not depend on the sense of the twist, i.e., the sign of $\delta$, and thus can be expanded by the square of $\delta$. The helicity modulus $\gamma$ is defined as the coefficient of the leading term of the expansion as,
\begin{equation}
F(\delta)-F(0)=\frac{\gamma}{2} \delta^2 + \mathcal{O}(\delta^4).
\label{eq:App01}
\end{equation}
Through transforming into the second-order differencial form and taking the limit of $\delta \rightarrow 0$, we see that $\gamma$ is given by the second-order derivative of the free energy $F$ with respect to $\delta$ as,
\begin{align}
\gamma
\approx 2\frac{F(\delta)-F(0)}{\delta^2}
=\frac{1}{\delta} \left(\frac{F(\delta)-F(0)}{\delta}
-\frac{F(0)-F(-\delta)}{\delta}\right)
\approx \left. \frac{\partial^2 F}{\partial \delta^2} \right |_{\delta=0}.
\label{eq:App02}
\end{align}
The three-dimensional spin vector $\bm S_i$ can be written with $S_i^z$ and $\phi_i$ as,
\begin{align}
\bm S_i
=\left(S_i^x, S_i^y, S_i^z \right)
=\left(\sqrt{1-{S_i^z}^2}\cos\phi_i, \sqrt{1-{S_i^z}^2}\sin\phi_i, S_i^z \right).
\label{eq:App03}
\end{align}
Using this expression, the distribution function $Z(\delta)$ of the twisted system is given by,
\begin{align}
Z(\delta)
=\int_{-2\pi}^{-2\pi} d\phi_1 \int_{-2\pi}^{-2\pi} d\phi_2 \cdots \int_{-2\pi}^{-2\pi} d\phi_{L^2} \; e^{-\beta\tilde{\mathcal{H}}(\delta)}
\equiv \int \mathcal{D}\phi \; e^{-\beta\tilde{\mathcal{H}}(\delta)}.
\label{eq:App04}
\end{align}
Here $\tilde{\mathcal{H}}(\delta)$ is the Hamiltonian of the twisted system. The helicity modulus $\gamma$ is calculated as,
\begin{align}
\gamma
&=\left. \frac{\partial^2 F}{\partial \delta^2} \right |_{\delta=0}
=\left. \frac{\partial^2}{\partial \delta^2} 
\left(-\frac{1}{\beta}\ln Z(\delta) \right) \right |_{\delta=0}
\nonumber \\
&=\left. \frac{\int\mathcal{D}\phi \;
\frac{\partial^2\tilde{\mathcal{H}}}{\partial\delta^2}
e^{-\beta\tilde{\mathcal{H}}}}{Z(\delta)} \right |_{\delta=0}
-\left. \beta\frac{\int\mathcal{D}\phi \; \left(
\frac{\partial\tilde{\mathcal{H}}}{\partial\delta} \right)^2
e^{-\beta\tilde{\mathcal{H}}}}{Z(\delta)} \right |_{\delta=0}
%
+\left. \beta\left( 
\frac{\int\mathcal{D}\phi \; \frac{\partial\tilde{\mathcal{H}}}{\partial\delta}
e^{-\beta\tilde{\mathcal{H}}} }{Z(\delta)} 
\right)^2 \right |_{\delta=0}
\nonumber \\
&=\left. 
\left\langle \frac{\partial^2\tilde{\mathcal{H}}}{\partial\delta^2} \right\rangle 
\right |_{\delta=0}
-\left. 
\beta\left\langle \left(\frac{\partial\tilde{\mathcal{H}}}{\partial\delta}\right)^2 
\right\rangle \right |_{\delta=0}
+\left. 
\beta {\left\langle \frac{\partial\tilde{\mathcal{H}}}{\partial\delta} \right\rangle}^2 
\right |_{\delta=0}.
\label{eq:App05}
\end{align}
When the system contains only symmetric exchange interactions associated with $\bm S_i \cdot \bm S_j$, the following relation holds,
\begin{equation}
\left. \left\langle \frac{\partial\tilde{\mathcal{H}}}{\partial \delta} \right\rangle \right |_{\delta=0} 
= \left. \left\langle \frac{\partial F}{\partial \delta} \right\rangle \right |_{\delta=0} =0.
\label{eq:App06}
\end{equation}
Consequently, the last term in Eq.~(\ref{eq:App05}) vanishes, and we obtain the following expression for a general form of the helicity modulus,
\begin{align}
\gamma=
\left. 
\left\langle \frac{\partial^2\tilde{\mathcal{H}}}{\partial\delta^2} \right\rangle 
\right |_{\delta=0}
-\left. \beta\left\langle 
\left(\frac{\partial\tilde{\mathcal{H}}}{\partial\delta}\right)^2 \right\rangle 
\right |_{\delta=0}.
\label{eq:App07}
\end{align}

Now we discuss the derivation of the formula of the helicity modulus for the XXZ model. The Hamiltonian of the XXZ model can be written using $S_i^z$ and $\phi_i$ as,
\begin{align}
\mathcal{H}
&=-J\sum_{<i,j>} \left \{ \sqrt{1-{S_i^z}^2} \sqrt{1-{S_j^z}^2}
(\cos\phi_i \cos\phi_j + \sin\phi_i \sin\phi_j) 
+\Delta S_i^z S_j^z \right \} 
\nonumber \\
&=-J\sum_{<i,j>} \left \{ \sqrt{1-{S_i^z}^2} \sqrt{1-{S_j^z}^2} 
\cos(\phi_i - \phi_j) 
+\Delta S_i^z S_j^z \right \}.
\label{eq:App08}
\end{align}
We consider the twisting of spin alignment in the $x$ direction. Subsequently, the term in the Hamiltonian associated with the bonds along the $x$ axis is modified as,
\begin{equation}
\tilde{\mathcal{H}}_x(\delta)=-J \sum_{<i,j>_x} \left \{ \sqrt{1-{S_i^z}^2} \sqrt{1-{S_j^z}^2} \cos\left( \phi_i - \phi_j - \frac{\delta}{L} \right) + \Delta S_i^z S_j^z \right \}.
\label{eq:App09}
\end{equation}
Note that the term associated with the bonds along the $y$ axis does not change. For the above Hamiltonian, we obtain,
\begin{align}
\frac{\partial^2\tilde{\mathcal{H}}}{\partial \delta^2}
&=\frac{\partial^2\tilde{\mathcal{H}}_x}{\partial \delta^2} 
=\frac{J}{L^2} \sum_{<i,j>_x} \sqrt{1-{S_i^z}^2} \sqrt{1-{S_j^z}^2} \cos \left( \phi_i - \phi_j - \frac{\delta}{L} \right), 
\label{eq:App10}
\\
\frac{\partial\tilde{\mathcal{H}}}{\partial \delta} 
&=\frac{\partial\tilde{\mathcal{H}}_x}{\partial \delta} 
=\frac{J}{L} \sum_{<i,j>_x} \sqrt{1-{S_i^z}^2} \sqrt{1-{S_j^z}^2} \sin \left( \phi_i - \phi_j - \frac{\delta}{L} \right).
\label{eq:App11}
\end{align}
Eventually, we obtain the following expression of the helicity modulus for the XXZ model,
\begin{align}
\gamma_{\rm XXZ}
&=\left. \left\langle \frac{\partial^2\tilde{\mathcal{H}}}{\partial \delta^2} \right\rangle \right |_{\delta=0} 
-\beta \left. \left\langle \left( \frac{\partial\tilde{\mathcal{H}}}{\partial \delta} \right)^2 \right\rangle \right |_{\delta=0} 
\nonumber \\
&=\frac{J}{L^2} \left \langle \sum_{<i,j>_x} \sqrt{1-{S_i^z}^2} \sqrt{1-{S_j^z}^2} \cos(\phi_i - \phi_j) \right \rangle
-\frac{\beta J^2}{L^2} \left \langle \left( \sum_{<i,j>_x} \sqrt{1-{S_i^z}^2} \sqrt{1-{S_j^z}^2} \sin(\phi_i - \phi_j) \right)^2 \right \rangle.
\label{eq:App12}
\end{align}
Sustituting $S_i^z=0$ into this expression, we obtain the expression for the XY model as,
\begin{align}
\gamma_{\rm XY}
&=\frac{J}{L^2} \left \langle \sum_{<i,j>_x} \cos(\phi_i - \phi_j) \right \rangle
-\frac{\beta J^2}{L^2} \left \langle \left( \sum_{<i,j>_x} \sin(\phi_i - \phi_j) \right)^2 \right \rangle.
\label{eq:App13}
\end{align}
\end{widetext}

\section{Acknowledgement}
This work is supported by Japan Society for the Promotion of Science KAKENHI (Grant No.~20H00337, 22H05114), CREST, the Japan Science and Technology Agency (Grant No.~JPMJCR20T1), and a Waseda University Grant for Special Research Projects (Project No.~2023C-140, 2023E-026).

\color{black}

\end{document}